\documentclass[review,authoryear]{elsarticle}

\usepackage{amssymb}
\usepackage{adjustbox}
\usepackage{array}
\usepackage{natbib}

\def\book#1 #2 #3 #4 {#1, #2. #3. #4.}
\def\chap#1 #2 #3 #4 #5 #6 #7 {#1, #2. #3. In: #4 (Eds.), #5. #6, pp.\ #7.}
\def\paper#1 #2 #3 #4 #5 #6 {#1, #2. #3. #4\ #5, #6.}
\def\inpress#1 #2 #3 #4 {#1, #2. #3. #4, in press.}
\def\submitted#1 #2 #3 #4 {#1, #2. #3. #4, submitted.}
\def\inprep#1 #2 #3 #4 {#1, #2. #3. #4, in preparation.}
\def\thesis#1 #2 #3 #4 #5 {#1, #2. #3. Thesis, #4. #5 pp.}
\def\abs#1 #2 #3 #4 #5 #6 {#1, #2. #3. #4, #5, pp.\ #6.}

\newcommand{\etal}{et al.}

\journal{Icarus}

\begin{document}

\begin{frontmatter}

\title{Constraining the final merger of contact binary (486958) Arrokoth with soft-sphere discrete element simulations}

\author{J. C. Marohnic\corref{Marohnic}\fnref{UMD}}
\cortext[Marohnic]{Corresponding author}
\ead{marohnic@astro.umd.edu}

\author{D. C. Richardson\fnref{UMD}}
\author{W. B. McKinnon\fnref{WUSTL}}
\author{H. F. Agrusa\fnref{UMD}}
\author{J. V. DeMartini\fnref{UMD}}
\author{A. F. Cheng\fnref{JHU/APL}}
\author{S. A. Stern\fnref{SWRI}}
\author{C. B. Olkin\fnref{SWRI}}
\author{H. A. Weaver\fnref{JHU/APL}}
\author{J. R. Spencer\fnref{SWRI}}

\author{the New Horizons Science team}

\fntext[UMD]{Department of Astronomy, University of Maryland, College Park, MD 20742, USA}
\fntext[JHU/APL]{Johns Hopkins University Applied Physics Laboratory, Laurel, MD 20723, USA}
\fntext[WUSTL]{Department of Earth and Planetary Sciences and McDonnell Center for the Space Sciences, Washington University in St. Louis, St. Louis, MO 63130, USA}
\fntext[SWRI]{Southwest Research Institute, Boulder, CO 80302, USA}

\begin{abstract}
The New Horizons mission has returned stunning images of the bilobate Kuiper belt object (486958) Arrokoth. It is a contact binary, formed from two intact and relatively undisturbed predecessor objects joined by a narrow contact region. We use a version of pkdgrav, an N-body code that allows for soft-sphere collisions between particles, to model a variety of possible merger scenarios with the aim of constraining how Arrokoth may have evolved from two Kuiper belt objects into its current contact binary configuration. We find that the impact must have been quite slow ($\lesssim$ 5 m/s) and grazing (impact angles $\gtrsim$ 75$^{\circ}$) in order to leave intact lobes after the merger, in the case that both progenitor objects were rubble piles. A gentle contact between two bodies in a close synchronous orbit seems most plausible.
\end{abstract}


\end{frontmatter}


\section{Introduction}
\label{S:1}

\subsection{(486958) Arrokoth}

\begin{figure}[h!]
\centering\includegraphics[width=\linewidth]{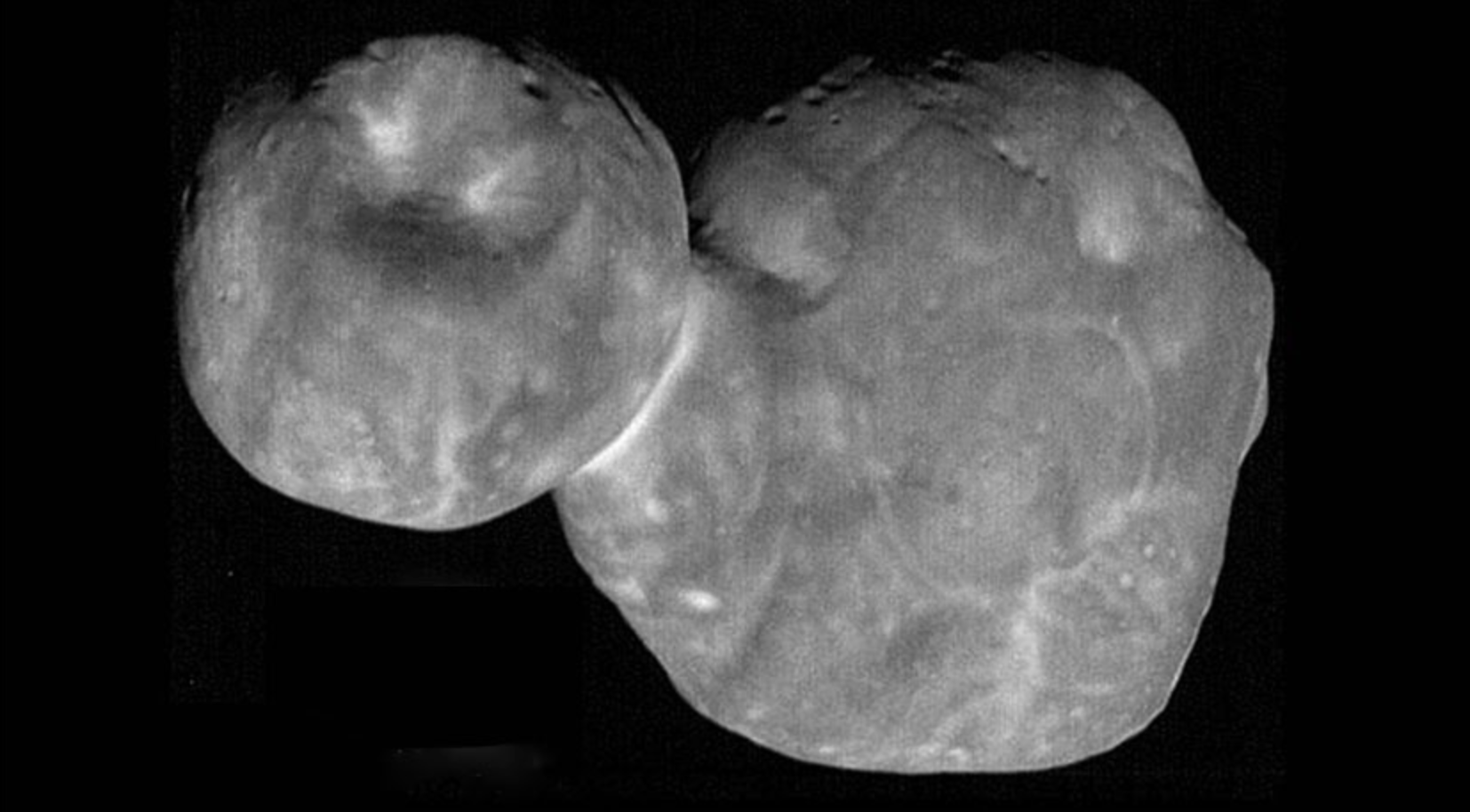}
\caption{An image of Arrokoth captured by New Horizons during its flyby. (Credit NASA, JHU APL, Southwest Research Institute, ESA)}
\label{fig:UT_image}
\end{figure}

After exploring the Pluto system in 2015, the New Horizons mission was redirected to include a visit to the Kuiper belt object (486958) Arrokoth. On January 1, 2019, the New Horizons spacecraft passed within 3550 kilometers of the body (Stern \etal, 2019). Arrokoth lies in the cold classical Kuiper belt and is a bilobate contact binary---we will refer to the larger of the two lobes as ``LL" and to the smaller as ``SL," following McKinnon \etal\ (2020) (see Figure \ref{fig:UT_image}). LL and SL have measured diameters of 19.5 km and 14.2 km, respectively. Both lobes approximate flattened or oblate spheres in shape. They are joined by a narrow contact region, which we will refer to as the ``neck." As is typical for a cold classical Kuiper Belt Object (CCKBO) (Gulbis \etal, 2006; Noll \etal, 2008), Arrokoth is red in color (Benecchi \etal, 2019a), though the neck is slightly less red and noticeably brighter than most of the rest of the object. Its rotation period is 15.92 $ \pm $ 0.1 hours. Thus far, no evidence of satellites larger than $\sim$2 km in diameter has been found, nor has any sign of gas or dust in the vicinity of Arrokoth (Stern \etal, 2019). Given its heliocentric distance, its surface temperature is likely somewhere between 30 and 60 K (Prialnik \etal, 2008). Its orbit has semimajor axis, eccentricity, and inclination of 44.08 au, 0.035, and 2.4$^{\circ}$, respectively (Stern \etal, 2018; Benecchi \etal, 2019b). 

\subsection{The origins of Arrokoth}

As a CCKBO, Arrokoth is likely among the most primitive bodies in the solar system. These bodies are thought to have formed in situ from the solar nebula approximately 4.5 Gyr ago and remained nearly undisturbed since (Batygin \etal, 2011). Considering its unusual shape, it seems likely that the present-day Arrokoth was created by the merger of two progenitor Kuiper Belt Objects (KBOs), which could have comprised a binary system prior to the merger or could have collided by chance. The present configuration and characteristics of Arrokoth may be able to give insight into the formation of Arrokoth in particular and into the formation of KBOs and Kuiper belt binaries (KBBs) in general.

Numerous possible formation mechanisms for KBOs have been proposed, which can be divided into two broad categories. The first is hierarchical coagulation (HC), in which successive two-body collisions lead to the gradual accretion of large objects. Competition between mergers on one hand and fragmentation on the other determine the outcome and the time required to form KBOs. Studies by Stern and Kenyon, among others, conclude that in order to form the Kuiper belt as it exists today in $\lesssim 10^8$ years via HC, orbits in the primordial belt would need to have had $e \sim 10^{-4}$ and $i \sim 10^{-2}$ (Stern, 1996; Kenyon, 2002; Nesvorn\'{y} \etal, 2010). The question of how binary KBOs might form under an HC model has been approached from many directions. Three-body encounters could leave behind a binary when a third object is ejected, removing angular momentum from the system (Goldreich \etal, 2002). Two KBOs might enter an unstable orbit by entering each other's Hill spheres or via chaos-assisted temporary capture and transition to a stable orbit once dynamical friction from background planetesimals removes excess energy (Goldreich \etal, 2002; Astakhov \etal, 2005; Lee \etal, 2007; Schlichting and Sari, 2008). Additionally, two bodies might collide within the Hill sphere of a third body, dissipating enough energy to form a stable binary with a mass ratio near one (Weidenschilling, 2002), a common characteristic of KBBs (Noll \etal, 2007; Schlichting and Sari, 2008). Such collisions could also produce a binary with a much larger mass ratio that then undergoes a series of exchange reactions with other KBOs, leaving behind a much larger secondary (Funato \etal, 2004).

An alternative to HC is gravitational instability (GI), in which gravitationally bound clouds of particles form in the thick midplane of the disk or via the streaming instability and collapse into large objects on short timescales (Nesvorn\'{y} \etal, 2010). Any angular momentum this system has will encourage the formation of two or more bound objects of similar sizes. Forming binaries via GI also gives a natural explanation for the observed correlation in color of KBBs (Benecchi \etal, 2009), as binaries formed locally would have the same composition and thus the same color. More recently, Nesvorn\'{y} \etal\ (2019) showed that GI can reproduce the observed KBB inclination distribution, while HC models cannot. While the simulations presented in this study model only the final stage of the contact binary formation, understanding the origins of the progenitor bodies helps to put the merger in context.

In spite of the merger that likely formed Arrokoth, LL and SL individually appear to be quite intact. This suggests that the contact binary was created by a low-speed collision between two primordial CCKBOs. Given their relatively small sizes, great distance from the Sun, and the fact that they lie in the dynamically cold segment of the Kuiper belt, CCKBOs are unlikely to have undergone any significant processing after their initial formation. As such, they can provide us with a unique probe into conditions in the outer solar nebula, and Arrokoth in particular should be able to give us insight into the formation of KBBs and of planetary bodies in general (Stern \etal, 2019; McKinnon \etal, 2020).

\subsection{Objectives}

We detail and expand on the numerical simulations introduced in McKinnon \etal\ (2020) to match the morphology and dynamical state of Arrokoth following the merger of its contact binary components. We constrain the merger circumstances by running a suite of simulations with different initial conditions and comparing the outcomes with observations made by the New Horizons spacecraft. Specifically, we consider the overall shape and final spin state of Arrokoth. We apply these methods exclusively to the final stages of the merger and do not model the genesis of the binary system prior to the merger event. 

Prior work has used numerical models to study the formation of bilobate comets and asteroids via mergers and collisions. Jutzi and Asphaug (2015) used an SPH code to simulate collisions between relatively small objects (0.1-1 km) at or just above the mutual escape speed, producing some bilobate objects. Subsequent studies have focused on higher-speed impacts, at 200 m/s or above, of small impactors hitting much larger target bodies (Jutzi and Benz, 2017; Jutzi, 2019; Sugiura \etal, 2020), though Schwartz \etal\ (2018) considers impacts as slow as 20 m/s. Most of this work was done with objects like comet 67P/Churymov-Gerasimenko in mind. They do not attempt to produce anything like the narrow neck and intact lobes that we see in Arrokoth, and focus on catastrophic or sub-catastrophic collisions that produce far more disruption than must have been the case in the merger that produced the Arrokoth contact binary. Nesvorn\'{y} \etal\ (2010) considers the possibility that 67P began as a binary and collapsed. They simulate a variety of merger scenarios and produce some very gentle collisions, but focus primarily on the dynamics of the binary collapse and do not carefully examine the merger itself. While we consider the possibility of a collision between unbound bodies in the case of Arrokoth, we also include a variety of simulations in which Arrokoth forms as the result of a very gentle merger between two bound objects.

\section{Method}
\label{S:2}

\subsection{pkdgrav}
To model the merger of Arrokoth, we use pkdgrav, a parallel $N$-body tree code (Stadel, 2001) with an implementation of the soft-sphere discrete element method (SSDEM) for collisions between spherical particles (Schwartz \etal, 2012). The SSDEM model works by allowing particles to interpenetrate slightly as a proxy for surface deformation, with restoring forces implemented as damped springs with a user-adjustable spring constant. The model includes a detailed implementation of static, rolling, and twisting friction (Zhang \etal, 2017) along with interparticle cohesion (Zhang \etal, 2018).

Given that pkdgrav is a parallel tree code, it can simulate systems of hundreds of thousands of particles quickly. This allowed us to run a relatively large number of high-resolution simulations to explore the parameter space. Furthermore, its explicit treatment of particle contacts makes pkdgrav well-suited to this energy regime. We are aware of at least one other numerical approach to modeling the final stage of the formation of Arrokoth (Wandel \etal, 2019), but not enough detail has been published at the time of this writing to allow a meaningful comparison between these approaches.

When used in this study, cohesive forces were only applied between particles of the same progenitor body. In other words, a contact between a particle from LL and a particle from SL would be treated as cohesionless. This choice was motivated by some of our early simulations of the binary merger with cohesion included. After an initial contact between bodies, the size of the neck would continue to grow as particles near the contact point stuck together and pulled others along with them. Because we wanted to use cohesion to capture the effect of material strength, we judged this behavior to be unphysical and made this adjustment to our model. The effects of cohesion in our simulations are discussed in greater detail in Section \ref{S:3}.

\subsection{Acceleration mapping}
To assess the extent to which the final merger disturbs surface material, we measured the net accelerations felt by particles on the surface of LL and SL throughout their mutual approach and impact. Specifically, we tracked accelerations due exclusively to contact forces between particles and recorded the maximum such acceleration experienced up to each point in the simulation by each particle. The aim of recording contact accrelerations was to observe shocks caused by the collision of the two bodies and to determine if this might have left a signature on the surface of Arrokoth.

\subsection{Setup of LL and SL}

We model each lobe of the present contact binary system separately, under the assumption that they formed separately and merged at some point in the past. We generate spherical ``rubble piles" out of many smaller particles. To create the oblate spheroids used in some of our simulations, we begin with the spherical bodies and remove particles as needed to ``carve out" the appropriate shapes. Thus, all oblate progenitor objects in this study have volumes less than their corresponding spherical progenitor bodies. The specific parameters used to generate LL and SL respectively (size, shape, number of particles, particle density) depend on the impact case being modeled (discussed in the following section). In the case of spherical components,\footnote{When this study began, the components were thought to be spherical. Although we now know they are oblate, the spherical case is useful for comparison purposes. See Section 2.4.} we use approximately 135,000 and 63,000 particles to model LL and SL, respectively. In our simulations, we adopted spherical radii of $R_{LL}$ = 8.97 km for LL and $R_{SL}$ = 6.82 km for SL. Because composing a body from equal-size particles adds excess shear strength that a real rubble pile would not likely have, we use a distribution of particle sizes (Quillen \etal, 2016; DeMartini \etal, 2019). Particle radii are normally distributed with a mean radius of 136 m and a standard deviation of 27 m. The largest and smallest radii included are 163 m and 109 m, respectively. Particles in a given simulation have a uniform mass density, though different densities are used in different runs. Most of the simulations presented here use a bulk density of 0.5 g/cm\textsuperscript{3}, though we include some simulations with lower densities (see Table \ref{table:setup}). Note that we measure the mass density of particles in a given simulation to be approximately twice the bulk density, due to the macroporosity of the rubble piles. In simulations with oblate lobes, the number of particles is in proportion to the decrease in volume from the original spherical shape. After generating LL and SL we run simulations with each body separately to allow the particles to settle into an equilibrium between self-gravity  and repulsive contact forces. 

Particles are either fricitionless or given ``gravel" friction parameters---with static, rolling, and twisting coefficients of friction of 1.0, 1.05, and 1.3, respectively. These parameters are representative of a typical rough material surface, and can be considered the opposite of the frictionless case. They correspond to material with a friction angle of approximately 39$^{\circ}$ for a random, polydisperse packing of spheres, similar to coarse, rocky materials on Earth (Zhang \etal, 2017). It should be noted here that the exact interior composition of Arrokoth is not known, although it is likely some combination of icy and rocky material (Stern \etal, 2019). The friction model in pkdgrav is largely designed to capture the effects of particle shape and internal friction rather than composition. On the assumption that Arrokoth is made up of irregular particles, we chose friction parameters that correspond to material with a reasonably high internal friction. We use a timestep of 0.037 seconds (chosen to sample the soft-sphere springs adequately) and run each simulation for one million steps, or about 10.3 hours. We aim to keep soft-sphere particle overlaps to $\sim$1\% of the smallest particle radius, so we take into account expected particle collision speeds and pressures and set the contact force spring constant accordingly---the value used in the simulations presented here is 8.0 x $10^{10}$ kg/s\textsuperscript{2}. We use a shape parameter of 0.5 and normal and tangential coefficients of restitution are both set to 0.2. The shape parameter describes how out-of-round the particles are treated for the purpose of computing friction and cohesion; smaller values indicate rounder particles and 0.5 has been found to be a good representation for gravel-like material. Our choice of restitution coefficients in this work results in more dissipation than for a canonical gravel-like material (with restitution coefficients of 0.55), but serves as a proxy for some crushing that is not otherwise captured in our code.\footnote{Estimates of the compressive strength of comet 67P/Churyumov-Gerasimenko are about 1 kPa (Heinisch \etal, 2018). A rough calculation suggests that the impact pressure felt by the two components of Arrokoth colliding at 5 m/s would be several orders of magnitude larger than this. On the other hand, the pressure cannot be so large that it visually distorts the bodies at the point of contact, as this is not evident in observations of Arrokoth. Thus, we can reasonably expect a moderate, but not excessive, amount of crushing.} Consequently, our results have the caveat that there may be more or less dissipation than is modeled here, which could alter the critical impact speed somewhat (Leinhardt \etal, 2000). The increased dissipation likely favors less body shape deformation at impact and a narrower neck. See Zhang \etal\ (2017) for more discussion of the parameters described in this section and their implementation in pkdgrav. 

\begin{table}[h!]
\vskip-2.9cm
\hskip-3cm
\begin{adjustbox}{angle=90}
\setlength\extrarowheight{-3pt}
\begin{tabular}{l c c c c c c c c}
\hline
& \textbf{Impact speed (m/s)} & \textbf{Impact angle} & \textbf{Friction} &\textbf{$\rho$ 
(g/cm\textsuperscript{3})} & \textbf{Cohesion (Pa)} & \textbf{$c_{LL}$} & \textbf{$c_{SL}$} & \textbf{Spins} \\
\hline
Slow inspiral \\
\hline
Inspiral 1 & 2.9 & 80$^{\circ}$ & Gravel & 0.5 & 0 & 1 & 1 & Synchronous \\ 
Inspiral 2 & 2.9 & 80$^{\circ}$ & Gravel & 0.5 & 275 & 1 & 1 & Synchronous \\ 
Inspiral 3 & 2.9 & 80$^{\circ}$ & Gravel & 0.5 & 2750 & 1 & 1 & Synchronous \\ 
Inspiral 4 & 2.9 & 80$^{\circ}$ & Gravel & 0.5 & 27500 & 1 & 1 & Synchronous \\ 
Inspiral 5 & 2.1 & 80$^{\circ}$ & Gravel & 0.25 & 0 & 1 & 1 & Synchronous \\ 
Inspiral 6 & 2.1 & 80$^{\circ}$ & Gravel & 0.25 & 275 & 1 & 1 & Synchronous \\ 
Inspiral 7 & 2.1 & 80$^{\circ}$ & Gravel & 0.25 & 2750 & 1 & 1 & Synchronous \\ 
Inspiral 8 & 2.1 & 80$^{\circ}$ & Gravel & 0.25 & 27500 & 1 & 1 & Synchronous \\ 
Inspiral 9 & 1.7 & 80$^{\circ}$ & Gravel & 0.16 & 0 & 1 & 1 & Synchronous \\ 
Inspiral 10 & 1.7 & 80$^{\circ}$ & Gravel & 0.16 & 275 & 1 & 1 & Synchronous \\ 
Inspiral 11 & 1.7 & 80$^{\circ}$ & Gravel & 0.16 & 2750 & 1 & 1 & Synchronous \\ 
Inspiral 12 & 1.7 & 80$^{\circ}$ & Gravel & 0.16 & 27500 & 1 & 1 & Synchronous \\ 
Inspiral 13 & 2.9 & 80$^{\circ}$ & None & 0.5 & 0 & 1 & 1 & Synchronous \\ 
Inspiral 14 & 2.9 & 80$^{\circ}$ & Gravel & 0.5 & 0 & 1 & 1 & None \\ 
Inspiral 15 & 2.9 & 80$^{\circ}$ & Gravel & 0.5 & 1000 & 1 & 1 & Equal, opposite\\ 
\hline
Oblate \\
\hline
Oblate 1 & 2.5 & 80$^{\circ}$ & Gravel & 0.5 & 0 & 1.5 & 1 & Synchronous \\ 
Oblate 2 & 2.5 & 80$^{\circ}$ & Gravel & 0.5 & 0 & 2 & 1.2 & Synchronous \\ 
Oblate 3 & 2.5 & 80$^{\circ}$ & Gravel & 0.5 & 0 & 2 & 1 & Synchronous \\ 
Oblate 4 & 2.5 & 80$^{\circ}$ & Gravel & 0.5 & 0 & 2 & 2 & Synchronous \\ 
Oblate 5 & 2.5 & 80$^{\circ}$ & Gravel & 0.5 & 0 & 3 & 1.4 & Synchronous \\ 
\hline
Impacts \\
\hline
Impact 1 & 5.0 & 45$^{\circ}$ & Gravel & 0.5 & 1000 & 1 & 1 & None \\ 
Impact 2 & 5.0 & 65$^{\circ}$ & Gravel & 0.5 & 1000 & 1 & 1 & None \\ 
Impact 3 & 5.0 & 75$^{\circ}$ & Gravel & 0.5 & 1000 & 1 & 1 & None \\ 
Impact 4 & 5.0 & 85$^{\circ}$ & Gravel & 0.5 & 1000 & 1 & 1 & None \\ 
Impact 5 & 10.0 & 45$^{\circ}$ & Gravel & 0.5 & 1000 & 1 & 1 & None \\ 
Impact 6 & 10.0 & 65$^{\circ}$ & Gravel & 0.5 & 1000 & 1 & 1 & None \\ 
Impact 7 & 10.0 & 75$^{\circ}$ & Gravel & 0.5 & 1000 & 1 & 1 & None \\ 
Impact 8 & 10.0 & 85$^{\circ}$ & Gravel & 0.5 & 1000 & 1 & 1 & None \\ 
Impact 9 & 1.0 & 45$^{\circ}$ & None & 0.5 & 0 & 1 & 1 & None \\ 
Impact 10 & 2.0 & 75$^{\circ}$ & None & 0.5 & 0 & 1 & 1 & None \\ 
\hline
\end{tabular}
\end{adjustbox}
\caption{A summary of the simulations performed. $\rho$ is the mass density and $c_{LL}$ and $c_{SL}$ are ratios of the long axes to the short (flattened) axes of LL and SL, respectively. Values listed in the sixth column are all \textit{interparticle} cohesive strengths.}
\label{table:setup}
\end{table}

\clearpage

\subsection{Impacts and merger events simulated}

As the initial returns from New Horizons suggested a contact binary composed of two \textit{spherical} lobes, we began with a suite of runs modeling mergers of two spherical bodies. While the shapes of LL and SL are now understood to be more similar to oblate spheroids, the spherical cases still serve as useful end members of a distribution of possible shapes and illustrate the effects of different parameters on the outcome of the merger event. Further, we include a number of scenarios in which the progenitor bodies are modeled as oblate spheroids, with one simulation matching the best-available shape model at the time of writing. While the actual shapes of LL and SL are not perfect spheroids, for the sake of simplicity we restrict ourselves to symmetric progenitors. Our range of flattened shapes was informed by the uncertainty measure of the observations. As this study was conducted, the best estimates of the shapes were evolving, and this is reflected in our choice of progenitor shapes.

To better understand the merger scenarios that could have led to the creation of Arrokoth as it exists today, we carried out a series of simulations, varying impact angle and impact speed, progenitor material cohesion, shape, density, and spin. We divide our simulations into three general groups: ``slow inspiral", ``oblate", and ``impacts". Each group consists of a grid search in which we vary different parameters to determine their effects on the final contact binary. See Table \ref{table:setup} for a summary of simulations performed in this study.

The ``slow inspiral" group attempts to model the final stage of a close, tidally locked orbit decaying toward an impact. In runs Inspiral 1--12, we vary density and cohesion. Based on the measured bulk densities of cometary nuclei (Groussin \etal, 2019) (for example, comets Tempel 1 and Churyumov-Gerasimenko (Samarasinha \etal, 2009; Jorda \etal, 2016)), we use 0.5 g/cm\textsuperscript{3} as a nominal starting point. We also test values of 0.25 g/cm\textsuperscript{3} and 0.16 g/cm\textsuperscript{3}. Four different cohesion values are also tested. To obtain a ``base" cohesion value, we divide the gravitational force felt by a particle at the surface of the spherical LL progenitor by the cross-sectional area of a mean-radius particle. This is equivalent to approximately 275 Pa. With this calculation, we do not attempt to take into consideration any measured or theoretical value for the interparticle cohesion of a KBO, but simply to get an idea of the scale appropriate to the body. This is all that is really necessary here, since we test a broad range of cohesion values anyway. We simulate inspiral scenarios with cohesion values of 0, 1, 10, and 100 times this value. We repeat this set of tests for the three bulk densities listed above. It should be noted that the values quoted here are \textit{interparticle} cohesion values, and unless explicitly stated otherwise, all mentions of cohesion in this paper refer to interparticle cohesion. The equivalent bulk cohesion is roughly 100 times smaller than these values for the type of material used in this study (Zhang \etal, 2018). Also included in this group are three runs in which friction and body spin are varied for an inspiraling orbit (Inspiral 13--15).

The ``oblate" group also consists of two tidally locked bodies spiraling inward, but uses oblate spheroids of varying axis ratios. Density is held constant at 0.5 g/cm\textsuperscript{3} and all runs in this group are cohesionless. The minor axes of the spheroids are parallel during the approach. At the time of writing, Oblate 5 represents the best oblate spheroid approximation of the true shapes of LL and SL.

The ``impacts" group does not assume that LL and SL are gravitationally bound before the merger, but tests direct impacts, varying collision angle and speed. We define the ``impact angle" as the angle between the relative velocity vector and the line connecting the centers of the two bodies at contact. Impact 1--8 test impact speeds of 5 m/s and 10 m/s at angles of 45$^{\circ}$, 65$^{\circ}$, 75$^{\circ}$, and 85$^{\circ}$ and use a cohesive strength of 1 kPa. It also includes two frictionless, low-speed impacts that we use to evaluate the effects of material type (Impact 9--10).

To get an estimate of the impact speeds relevant to the problem, we estimate the mutual escape speed of LL and SL. If we make the assumption that LL and SL are spheres in contact at a single point, then their mutual escape speed can be derived by setting the two-point-mass total energy to zero:

\begin{equation}
\label{eq:1}
\frac{v_{esc}^2}{2} - \frac{GM}{R} = 0,
\end{equation}

where $v_{esc}$ is the relative velocity between LL and SL, $M$ is their combined total mass, and $R$ is the distance between their centers. Their escape speed is then given by

\begin{equation}
\label{eq:2}
v_{esc} = \sqrt{\frac{2GM}{R}}.
\end{equation}

\noindent For a bulk density of 0.5 g/cm\textsuperscript{3} and radii of $R_{LL}$ = 8.97 km and $R_{SL}$ = 6.82 km, this gives an escape speed of 4.3 m/s. The escape speed should be close to the minimum possible impact speed we might expect for progenitor bodies \textit{not} in orbit about one another (i.e., zero relative speed at infinite distance). This motivates us to test direct impacts at 5 m/s and 10 m/s. If the lobes cannot survive collisions near the mutual escape speed, formation by direct impact becomes unlikely.

In cases where we wanted to reproduce a gradual spiral-in process with an impact angle near 90$^{\circ}$ (those in the ``slow inspiral" and ``oblate" groups), we initiated the s2imulations with the bodies in an approximately circular orbit. To calculate the appropriate speeds, we start with Kepler's third law 

\begin{equation}
\label{eq:3}
P^2 = \frac{4\pi^2}{GM}a^3,
\end{equation}

\noindent where $P$ is the orbital period of the system, $a$ is the semimajor axis of the orbit, and $M$ is the total mass. Let $R_{LL}$, $R_{SL}$, $M_{LL}$, and $M_{SL}$ be the radii and masses of LL and SL. We give LL and SL an initial separation of $1.04(R_{LL} + R_{SL})$ (leaving the bodies slightly separated to avoid particle overlaps) and a total mass of $M = M_{LL} + M_{SL}$. We are interested in producing a grazing contact, but due to the irregularities of the surface, setting up scenarios precisely so that they result in a specific point of contact is difficult. To obtain a reasonable starting point, we calculate the orbital speed that the bodies would have if they were in contact at a single point. To find the period for our system, we solve for $P$ and substitute:

\begin{equation}
\label{eq:4}
P = \sqrt{\frac{4\pi^2}{G(M_{LL} + M_{SL})}} \hspace{3pt}(R_{LL} + R_{SL})^{3/2}
\end{equation}

\noindent The orbital speed for each body is then given by

\begin{equation}
\label{eq:5}
v_{orb} = \frac{2\pi r_{orb}}{P},
\end{equation}

\noindent where $r_{orb}$ is the orbital radius of each body (4.9 km for LL and 11.5 km for SL, relative to their mutual center of mass). This gives a $v_{orb}$ of 0.9 m/s for LL and 2.0 m/s for SL in the case of Inspiral 1. In order to produce a collision in a reasonable time span, we impart both LL and SL with 90\% of the $v_{orb}$ values calculated in the manner described above. We emphasize that these initial speeds will not result in a circular orbit with the bodies in contact at precisely one point, but a slightly elliptical trajectory (representing the end-stage of an inspiral) that leads to a collision at nearly $90^{\circ}$. In any case, any departures from sphericity in the progenitor objects will mean that these calculations overestimate their speeds. As we are interested in the collisions and not the orbits themselves, this difference is unimportant.

\subsection{Neck measurements}
In order to compare the results of our simulations to each other and the observed system, we devised a method for computing the neck widths of the contact binaries produced by our simulations. We divide the longest axis of the body into equal-length segments. All particles falling inside a segment are grouped together, creating cross-sectional ``slices" of the object. For each slice, we calculate the maximum perpendicular distance between a particle in that slice and the central axis of the body. This allows us to get an estimate of the width of the object at a given point along the body's long axis, even when the object is not symmetrical. To calculate the width of the neck, we find the minimum width near the middle of the long axis. We report these measured values in Table \ref{table:outcomes}. For severely misshapen bodies, this method does not give a meaningful result, but for contact binaries similar in shape to Arrokoth, it gives a good quantitative estimate of the neck width. We report neck widths as a fraction of Arrokoth's observed long axis length. We use the value 0.184 for the physical Arrokoth as a point of comparison, though the exact figure depends on how one measures since the neck is not symmetric.

\section{Results}
\label{S:3}

Results of the simulations are summarized in Table \ref{table:outcomes}, showing the run label, final spin period, neck width, and a brief description of the final state of the lobes. Entries with ``N/A" for the period and neck width were cases that did not result in a contact binary. In the remainder of this section, we provide more detail on the results of the major parameters we tested, namely friction, bulk density, cohesion, initial spin, component shape, and impact speed and angle. A discussion of these results, which are summarized in Figure \ref{fig:widths}, is given in Section \ref{S:4}.

\begin{table}[htbp]
\hskip-1cm
\setlength\extrarowheight{-3pt}
\begin{tabular}{l c c c}
\hline
& \textbf{Final spin period (h)} & \textbf{Neck width} & \textbf{Final state of lobes}\\
\hline
Slow inspiral \\
\hline
Inspiral 1 & 9.0 & 0.184 & Intact\\ 
Inspiral 2 & 9.0 & 0.179 & Intact\\ 
Inspiral 3 & 8.9 & 0.184 & Intact\\ 
Inspiral 4 & 8.6 & 0.225 & Slight deformation\\ 
Inspiral 5 & 12.9 & 0.179 & Intact\\ 
Inspiral 6 & 12.9 & 0.179 & Intact\\ 
Inspiral 7 & 12.7 & 0.184 & Intact\\ 
Inspiral 8 & 12.3 & 0.276 & Moderate deformation\\ 
Inspiral 9 & 16.1 & 0.189 & Intact\\ 
Inspiral 10 & 16.1 & 0.189 & Intact\\ 
Inspiral 11 & 15.8 & 0.194 & Intact\\ 
Inspiral 12 & 15.9 & 0.292 & Significant deformation\\ 
Inspiral 13 & 8.9 & 0.332 & Significant deformation\\ 
Inspiral 14 & 11.6 & 0.317 & Slight deformation\\ 
Inspiral 15 & 9.6 & 0.297 & Slight deformation\\ 
\hline
Oblate \\
\hline
Oblate 1 & 9.7 & 0.174 & Intact\\ 
Oblate 2 & 10.7 & 0.169 & Intact\\ 
Oblate 3 & 10.4 & 0.169 & Intact\\ 
Oblate 4 & 11.5 & 0.179 & Intact\\ 
Oblate 5 & 11.7 & 0.205 & Intact\\ 
\hline
Impacts \\
\hline
Impact 1 & 9.0 & 0.358 & Significant deformation\\ 
Impact 2 & N/A & N/A & Moderate deformation \\ 
Impact 3 & N/A & N/A & Slight deformation \\ 
Impact 4 & N/A & N/A & Intact \\ 
Impact 5 & N/A & N/A & Significant deformation \\ 
Impact 6 & N/A & N/A & Moderate deformation \\ 
Impact 7 & N/A & N/A & Slight deformation \\ 
Impact 8 & N/A & N/A & Intact \\ 
Impact 9 & N/A & N/A & Completely merged \\ 
Impact 10 & N/A & N/A & Completely merged \\ 
\hline
\end{tabular}
\caption{A summary of the outcomes of the simulations. Final spin period and neck width are listed only if a recognizable contact binary results from the collision. Neck width is given as a fraction of Arrokoth's observed long axis length (35 km)}
\label{table:outcomes}
\end{table}

\subsection{Effects of friction}
\begin{figure}[h!]
\centering\includegraphics[width=\linewidth]{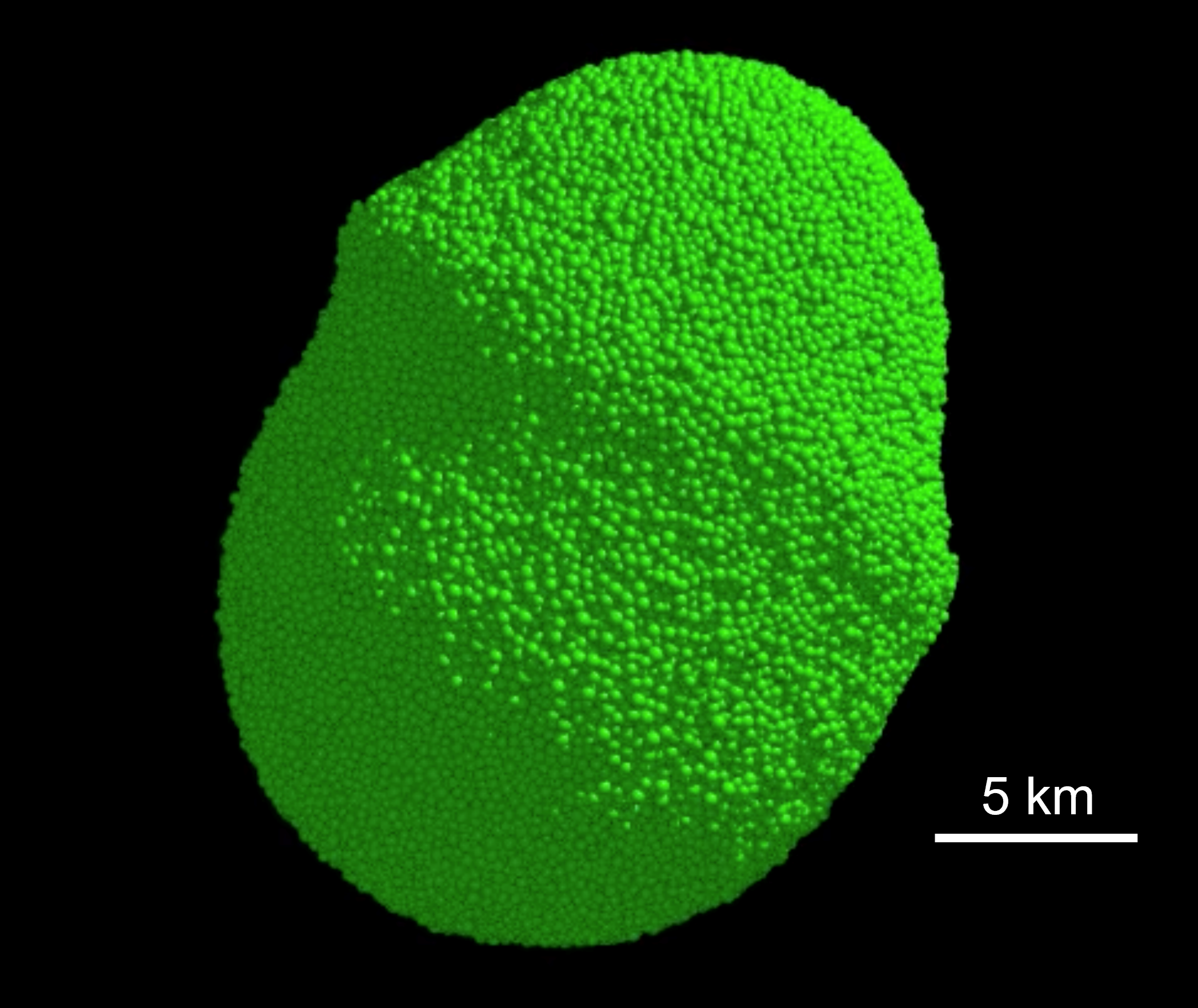}
\caption{Impact 9. 45$^{\circ}$ impact angle, 1 m/s, frictionless. LL and SL merge almost completely, leaving little evidence that the object was once a binary.}
\label{fig:b45v100_nofric}
\end{figure}

Impact 9, Impact 10, and Inspiral 13 are frictionless, while all other runs use the ``gravel" friction parameters described in Section \ref{S:2}. Impact 9 and Impact 10 both result in the total loss of the bilobate shape after contact merger---the progenitor bodies deform smoothly into one large mass (see Figure \ref{fig:b45v100_nofric}). Inspiral 13 preserves the shapes of LL and SL, but still produces a very wide neck. We conclude that material friction is important in maintaining the overall shape and neck width of the final shape of the contact binary.

\subsection{Density, cohesion, and spin}

\begin{figure}[h!]
\centering\includegraphics[width=\linewidth]{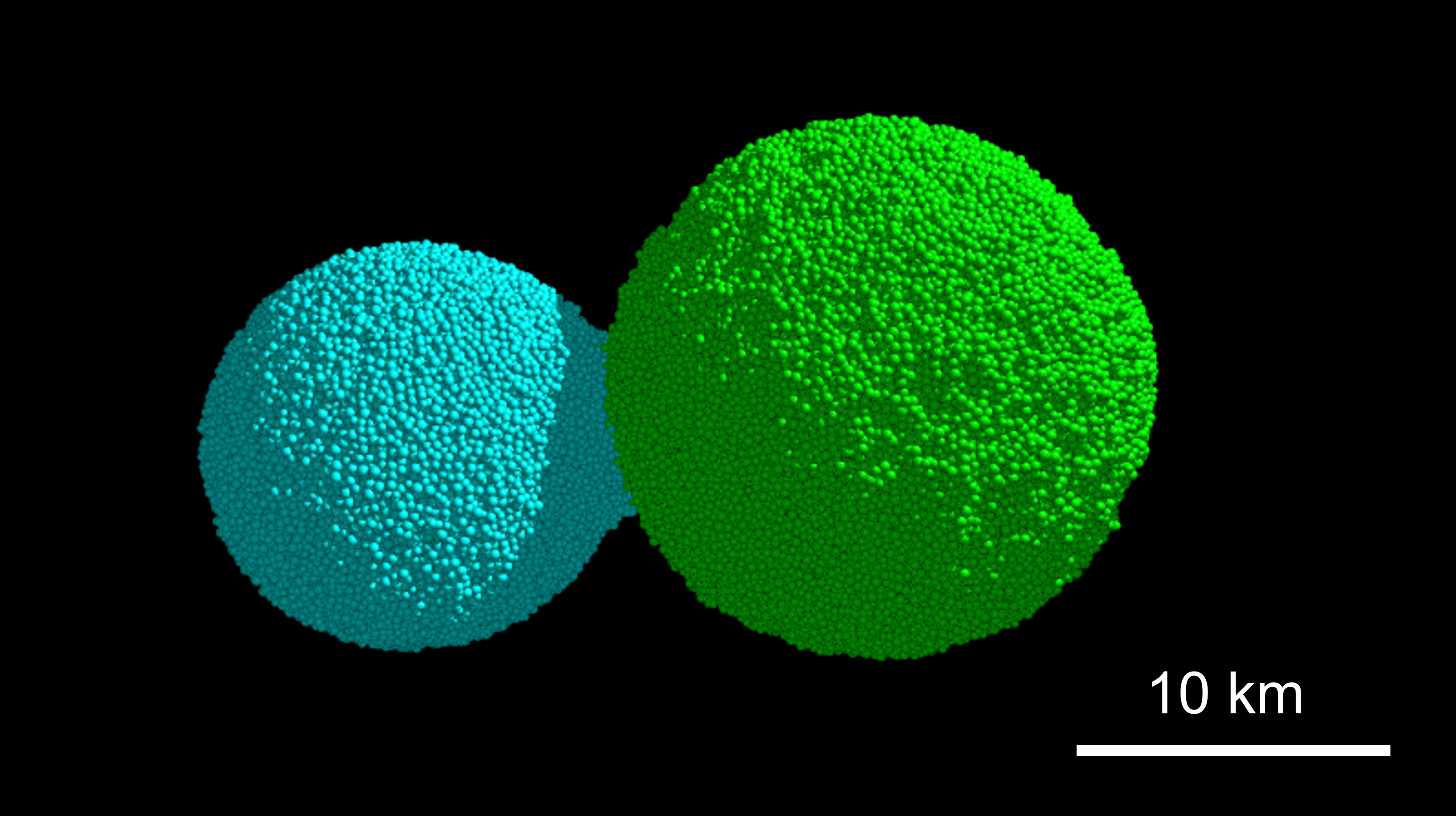}
\caption{Inspiral 1. Cohesionless with gravel friction parameters. Both lobes are left intact and the contact between them forms a well-defined, narrow neck.}
\label{fig:nominal}
\end{figure}

\begin{figure}[h!]
\centering\includegraphics[width=\linewidth]{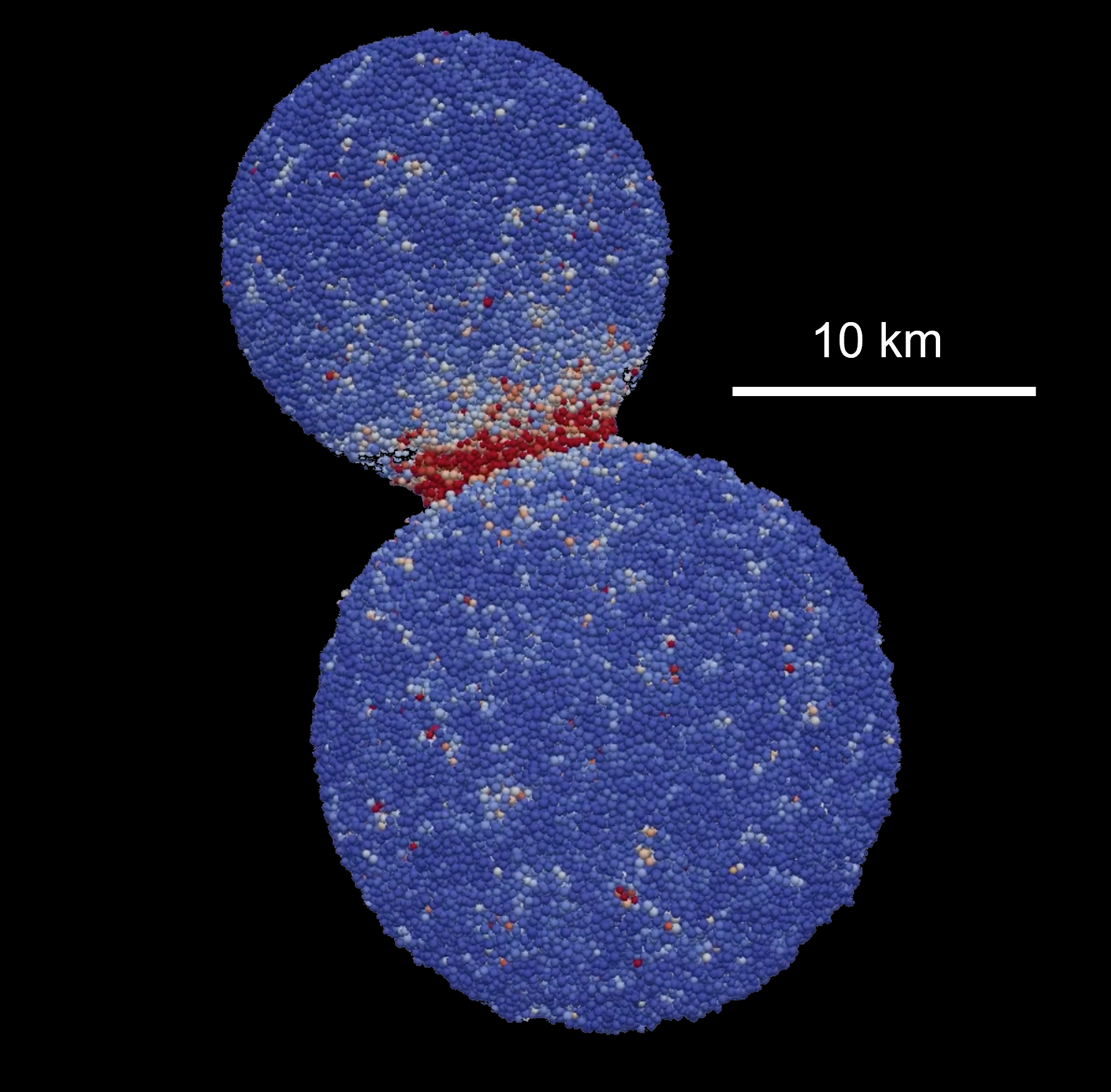}
\caption{Map of maximum total accelerations experienced by particles in Inspiral 1. Darkest reds correspond to 8.8 x $10^{-1}$ m/s\textsuperscript{2} and darkest blues correspond to 3.5 x $10^{-4}$ m/s\textsuperscript{2} on a linear scale. For comparison, the surface gravity of a sphere with volume equal to that of Arrokoth is $1.7 \times 10^{-3}$ m/s\textsuperscript{2}.}
\label{fig:nominal_acc}
\end{figure}

Inspiral 1--12 vary bulk density between 0.16 g/cm\textsuperscript{3} and 0.5 g/cm\textsuperscript{3} and interparticle cohesion between 0 and 27.5 kPa. At 0.5 g/cm\textsuperscript{3} (Inspiral 1--4), varying the cohesion does not have a strong effect on the outcome. Final spins range from 8.6 h to 9.0 h and neck widths vary between 0.179 and 0.225. We do notice, however, that at the highest cohesion value tested, we see somewhat more deformation, a slight increase in neck width, and a slightly shorter spin period (see Section \ref{S:4} for further discussion of this trend). Acceleration maps show disruption confined to the neck area (see Figures \ref{fig:nominal} and \ref{fig:nominal_acc}).

\begin{figure}[h!]
\centering\includegraphics[width=\linewidth]{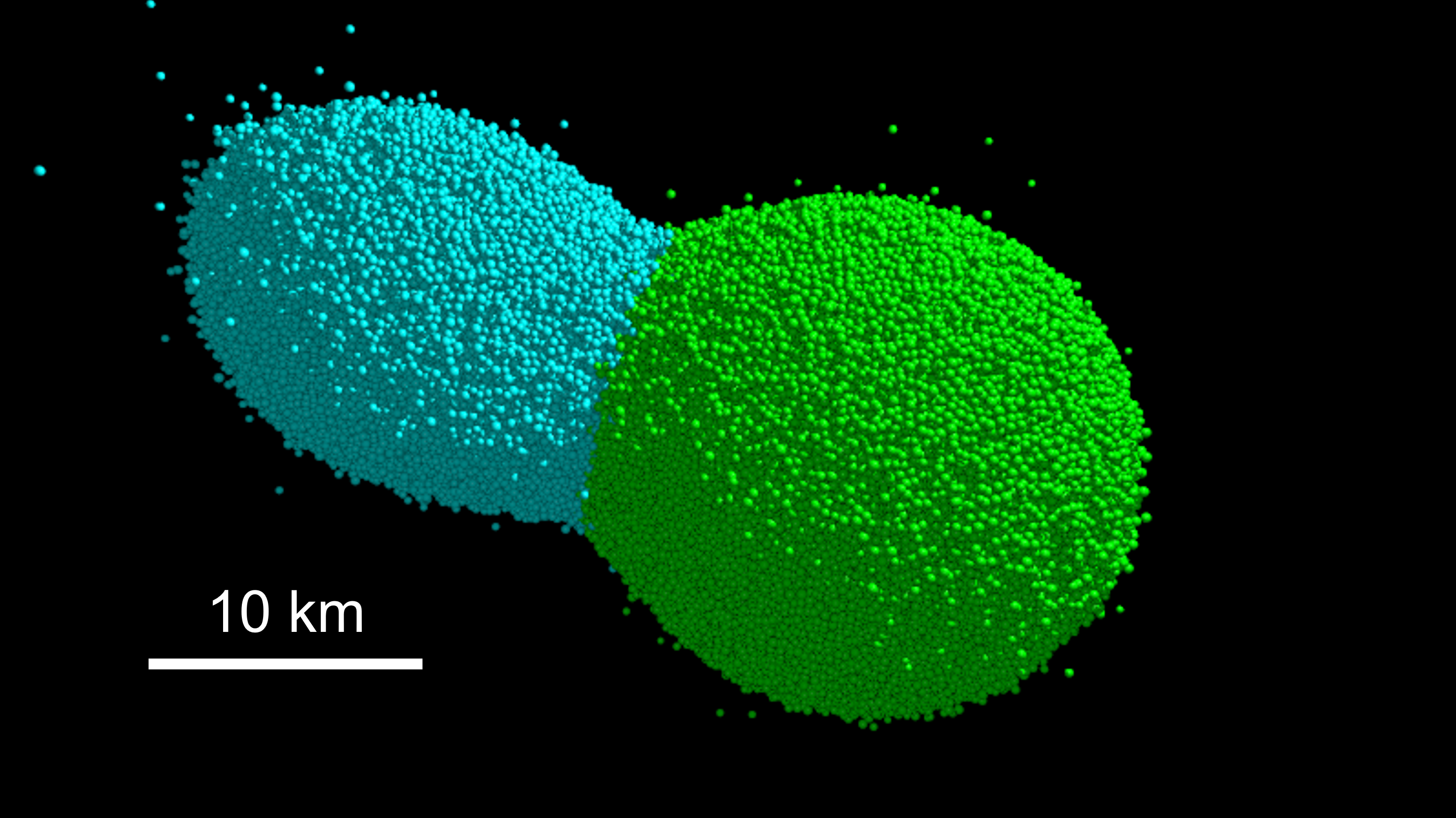}
\caption{Inspiral 12. Interparticle cohesion of 27500 Pa and gravel friction parameters. Suprisingly, low-density lobes (here 0.16 g/cm\textsuperscript{3}) with high interparticle cohesion result in more deformation, as the avalanche toward the neck tends to pull more material with it. The low-density bodies don't have enough self-gravity to hold their shapes.}
\label{fig:rho016coh100}
\end{figure}

\begin{figure}[h!]
\centering\includegraphics[width=\linewidth]{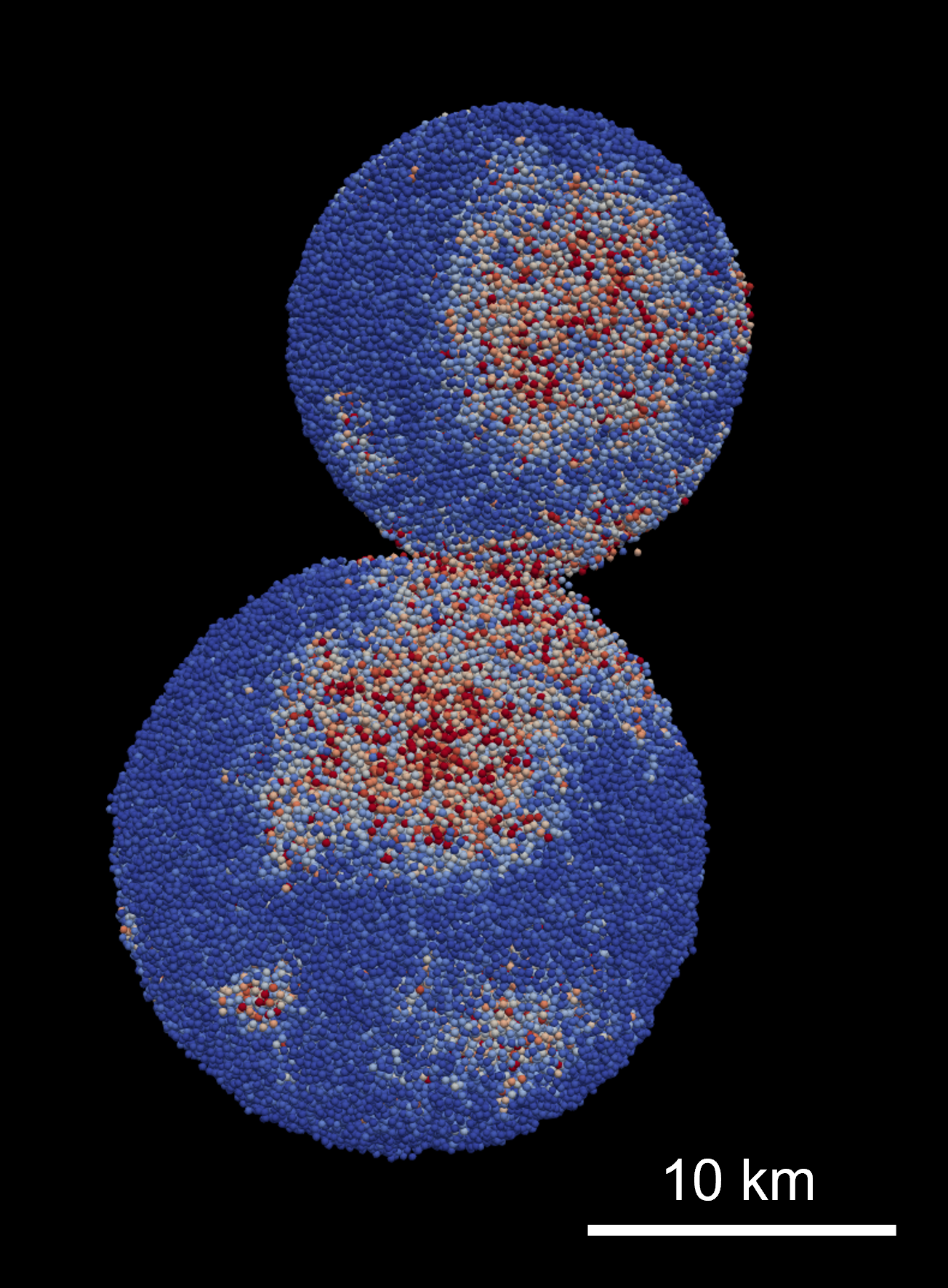}
\caption{Inspiral 12. Map of maximum contact accelerations. Darkest reds correspond to 2.0 m/s\textsuperscript{2} and darkest blues correspond to 0.0 m/s\textsuperscript{2} on a linear scale. Stresses are not localized to the neck as they are in Inspiral 1.}
\label{fig:rho016coh100_acc}
\end{figure}

At lower densities, this trend becomes more pronounced. For bulk densities of 0.25 g/cm\textsuperscript{3} (Inspiral 5--8), spin periods are 12.9 h and neck widths are 0.179 for the cohesionless and 275 Pa cases. At 27.5 kPa, the spin period falls to 12.3 h, the final neck width is 0.276, and the lobes deform noticeably. At 0.16 g/cm\textsuperscript{3} (Inspiral 9--12), the cohesionless case results in a spin period of 16.1 h and a neck width of 0.189, while the highest cohesion tested produced a spin of 15.9 h, a neck width of 0.292, and a severely deformed contact binary (see Figures \ref{fig:rho016coh100} and \ref{fig:rho016coh100_acc}). Generally, at a given density, increasing cohesion decreases spin period and increases neck width. This relation is much stronger at lower densities. Fixing cohesion and decreasing density similarly decreases spin period and increases neck width.

Finally, Inspiral 14 and Inspiral 15 vary the spin of the progenitor bodies. LL and SL have no spin in Inspiral 14 and in Inspiral 15, LL spins at the synchronous rate, while SL spins at the same rate but in the opposite direction. We find that when neither LL nor SL have spin or when they have opposite spins, the contact binary that they form is somewhat lopsided and has a relatively thick neck, with widths of 0.317 and 0.297, respectively. The non-rotating case produces a spin period of 11.6 h, while the counter-rotating case produces a spin period of 9.6 h. 

\subsection{Progenitor body shape}

In runs Oblate 1--5, the impacting components have oblate rather than spherical shapes. All simulations in this group model slowly inspiraling orbits with bulk densities of 0.5 g/cm\textsuperscript{3}, and synchronous rotations---characteristics we judged most likely to produce something similar to Arrokoth. Oblate 1--5 all produce relatively thin necks. Oblate 5 is the thickest at 0.205 and Oblate 2 and Oblate 3 are the thinnest at 0.169, pointing to no clear relation between body shape and neck width in these cases. Oblate 1 yields the shortest final period at 9.7 h, and Oblate 5 the longest at 11.7 h. Final spin generally decreases with increasing oblateness, though this is more directly a function of total progenitor mass, as described in Section \ref{S:4}. All oblate cases tested produce well-formed contact binaries, with clearly intact LL and SL lobes.

\subsection{Impact speed and angle}

\begin{figure}[h!]
\centering\includegraphics[width=\linewidth]{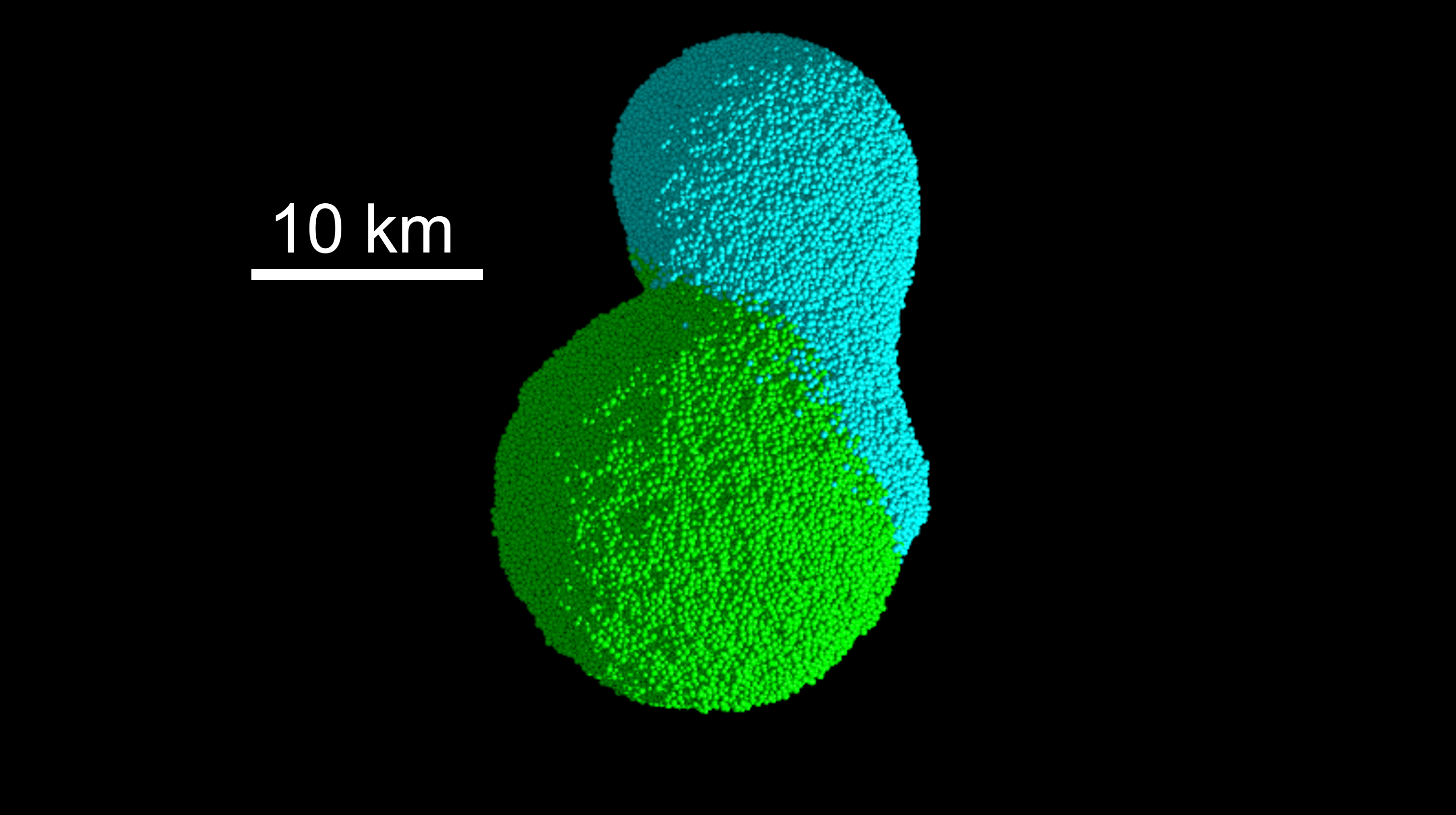}
\caption{Impact 1. 45$^{\circ}$ impact angle, 5 m/s. The impact creates a contact binary, but with an asymmetric neck and a lopsided SL lobe. The green and blue particles come from LL and SL, respectively.}
\label{fig:b45v500}
\end{figure}

\begin{figure}[h!]
\centering\includegraphics[width=\linewidth]{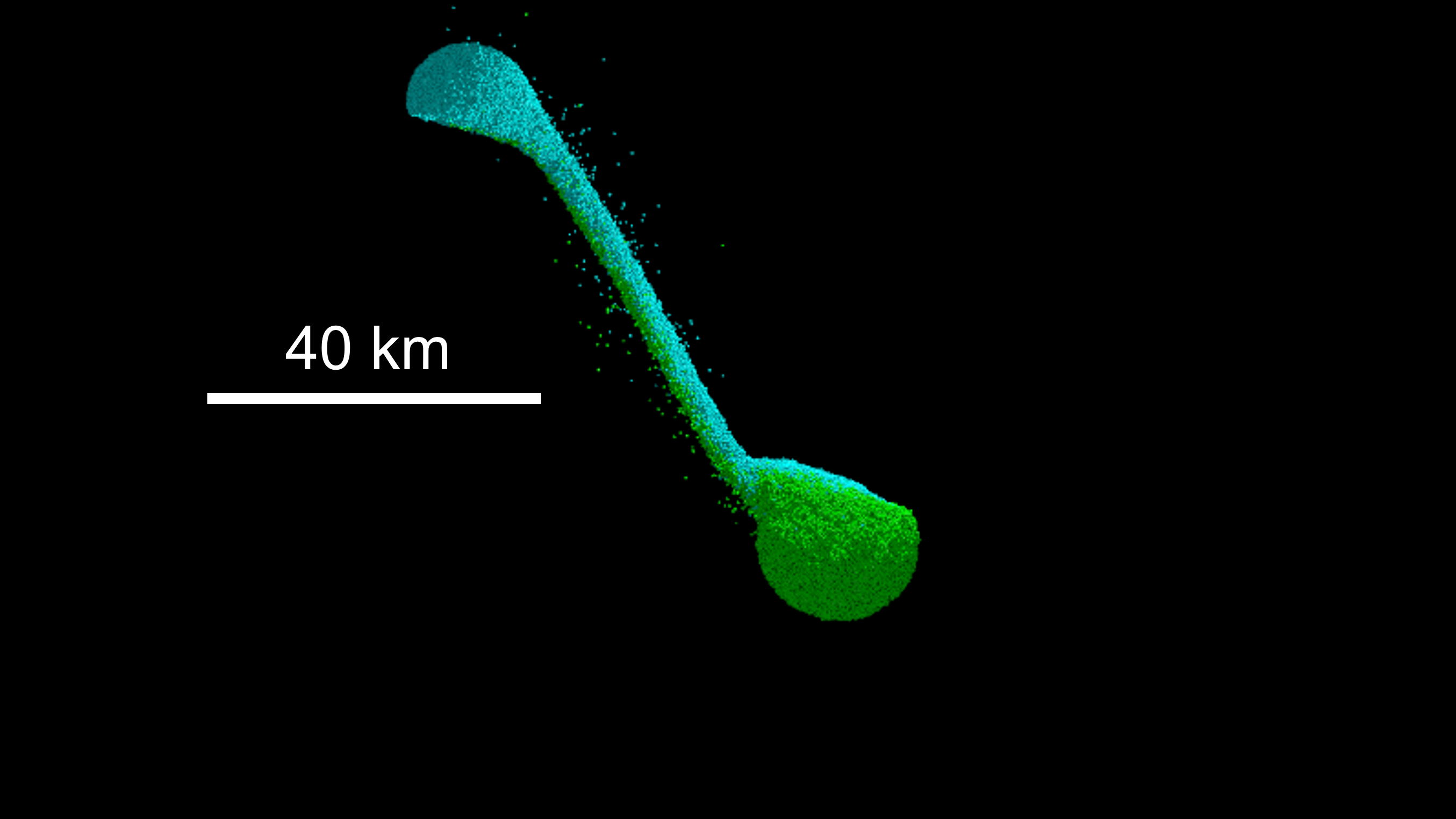}
\caption{Impact 5. 45$^{\circ}$ impact angle, 10 m/s. The impact severely disrupts both bodies, leaving a long tongue of material stretched between them. As the simulation progresses, this connection breaks as SL moves farther from LL.}
\label{fig:b45v1000}
\end{figure}

We test the effects of varying the speed and impact angle of direct collisions in Impact 1--8. Impact 1, at 5 m/s and 45$^{\circ}$, is slow and direct enough to produce a recognizably bilobate object, though it appears quite lopsided and its neck (width 0.358) is considerably thicker than that of the real Arrokoth (see Figure \ref{fig:b45v500}). None of the remaining cases in this group produce a contact binary. Impacts at $\geq 75^{\circ}$ lead to glancing contacts after which the progenitors separate mostly unscathed. Lower-angle impacts lead to moderate or even catastrophic disruption of the progenitor bodies (see Figure \ref{fig:b45v1000}). 

\section{Discussion}
\label{S:4}

\begin{figure}[h!]
\centering\includegraphics[width=\linewidth]{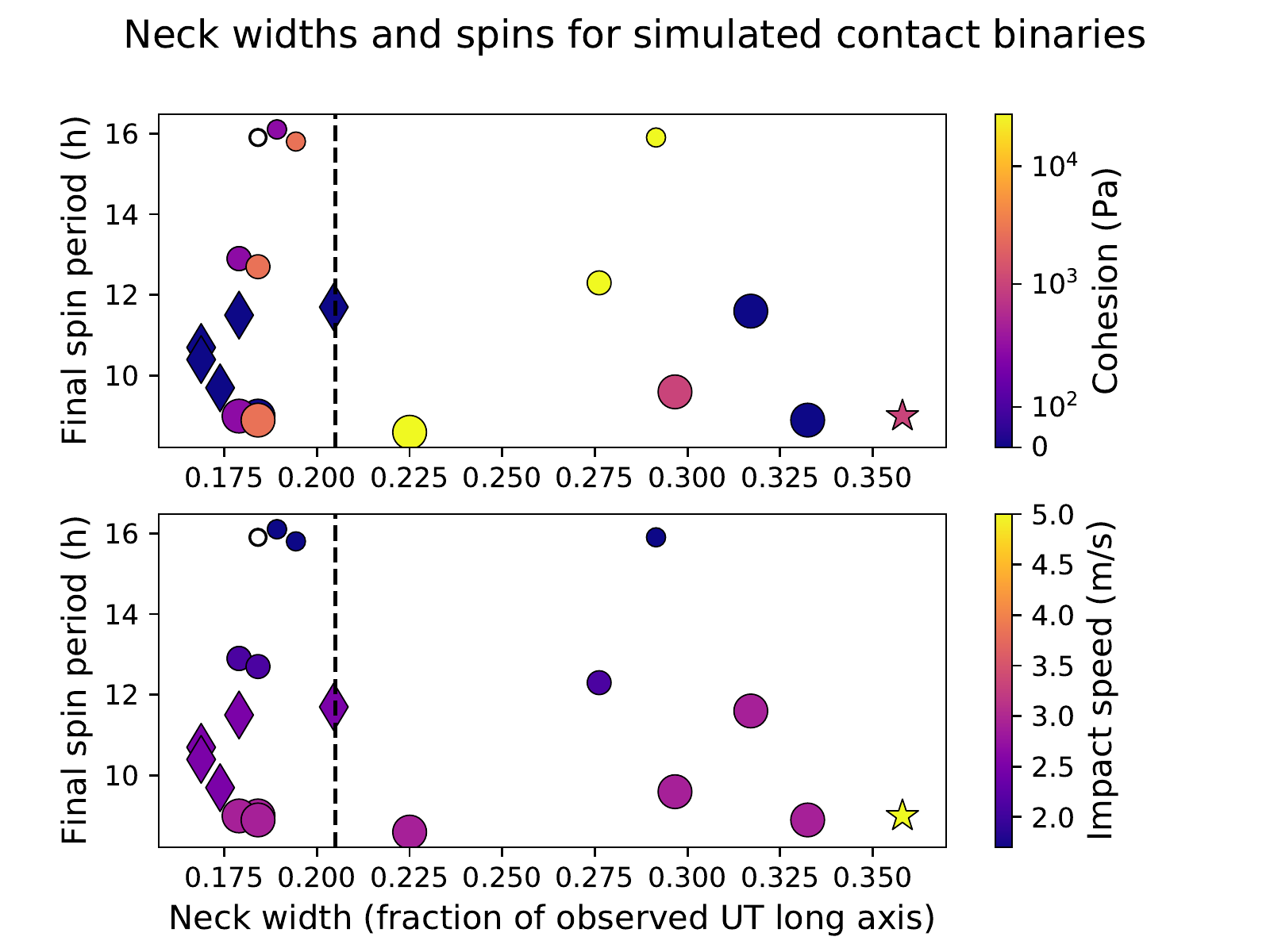}
\caption{Neck widths versus final spin periods for those simulations that produced a contact binary shape. Filled circles represent spherical inspiral runs, diamonds represent oblate inspiral runs, and stars represent direct impacts. The single open circle represents the observed values for Arrokoth. The vertical dashed line at 0.205 is taken as the likely division between plausible and implausible scenarios. Simulations lying to the right of this line produce neck widths too large to be a good match for Arrokoth. We derive this cutoff by considering neck widths within 10\% of Arrokoth's to be good matches. No runs produced neck widths that fell below the lower end of this range. Marker size is proportional to bulk density. Colors in panel 1 correspond to cohesion (logarithmic scale), while colors in panel 2 correspond to impact speed (linear scale).}
\label{fig:widths}
\end{figure}

Our aim in this study is to place constraints on the merger circumstances that led to the formation of the Kuiper belt contact binary Arrokoth. We want to determine the combination of parameters and impact circumstances that best reproduces the observed characteristics of Arrokoth, with intact lobes joined by a narrow neck and a spin period of 16 h. These are the criteria by which we judge the outcome of each simulation. However, we consider the spin period to be less important than the neck width and overall shape, as events after the merger may have modified the period. The Arrokoth contact binary may well have formed with a spin period much shorter than it has today and gradually spun down (through energy exchange with surrounding material, or even very slowly via the YORP effect---see McKinnon \etal\ (2020)). We also note that when we gradually increase the spin period of the contact binary produced by Inspiral 1 from 9 h to 16 h (using a spin-down procedure similar to that described by Zhang \etal\ (2017), the object retains its shape and neck width. This suggests that the period of the object immediately after the collision is somewhat incidental in judging whether a given merger scenario could have created Arrokoth. In Figure \ref{fig:widths} we plot the final neck widths and spin periods of all simulations that produce a contact binary, giving a graphical representation of how successful each simulation is in replicating the present-day Arrokoth. We use 0.205 as an upper limit for a plausible target neck width (shown by the vertical dashed line in the figure).

\subsection{Simulation outcomes}
Among the slow inspiral group, Inspiral 1, 5, 6, and 9 come closest to matching the observed neck width, while Inspiral 9, 10, 11, 12 best match Arrokoth's spin. The observed decrease in final spin with bulk density is a function of the total mass of the bodies. Because volume is kept constant across runs in this group, the lower density progenitors are less massive, and the speeds needed for them to maintain a synchronous orbit are lower. These lower orbital speeds translate to lower rotation speeds after the merger. As discussed above however, final spin period is a less important diagnostic than final shape, because the spin could change after formation (without necessarily changing the body shape).

We also see in the case of Inspiral 1--12 that high interparticle cohesion actually produces a thicker neck and a more distorted final shape. In the model used in this study, cohesion is only applied to particles in physical contact, and when it is active it introduces a force that may enhance or oppose gravity, depending on where the particles lie in the effective potential of the system. It appears that, due to the unusual shape of Arrokoth, high cohesion increases the tendency of surface material to flow toward the neck. When cohesive forces dominate over gravity, particles farther from the neck can be accelerated more violently, and those that detach may be moving quickly enough to temporarily leave the surface. This effect is more pronounced at lower bulk densities. In Inspiral 8 and, to an even greater extent, Inspiral 12 it causes disruption across the entire surface of the object after the merger (see Figures \ref{fig:rho016coh100} and \ref{fig:rho016coh100_acc}). We emphasize here that while our cohesion model produces this outcome, the physical cohesive forces at play on Arrokoth may not. Since the true frictional and cohesive properties of Arrokoth's constituent particles are unknown, we can only address the behavior of our model and a future study would be needed to examine other possible cohesion models.

Inspiral 13 demonstrates the importance of friction in maintaining the shapes of LL and SL, even when the centrifugal force from rotation aids in keeping the bodies separated. Finally, Inspiral 14 and 15 both test the effect of non-synchronous rotation, and both produce contact binaries that are misshapen with respect to Arrokoth. On the basis of final shape and neck width, we can likely rule out Inspiral 4, 8, 12, 13, 14, and 15. 

All runs in the oblate group produce contact binaries with intact lobes and neck widths in a plausible range. Because the bulk density is fixed while volume varies, the total mass is different in each run. As in the case of the inspiral group, this leads to a variation in final spin period due to differences in initial orbital speed. From the point of view of neck width and shape, all of these scenarios are plausible, though Oblate 4 and 5 are of course preferable as they represents collisions between two objects that most closely match the present-day shapes of LL and SL.

The impacts group is the least successful overall in terms of reproducing Arrokoth's spin and neck. Only Impact 1 results in a contact binary, and even this scenario can be ruled out as it does not match Arrokoth's shape. A grazing collision could leave the progenitor objects gravitationally bound and lead to a future collision. However such a collision would likely be at or near the mututal escape speed, which should create a deformed contact binary, unless the orbit first circularized and then gradually shrank, leading to a situation more akin to that described in the inspiral scenarios. Although impacts at 5 m/s are slightly above the estimated mutual escape speed of the progenitor objects, simulated impacts at 3.95 m/s (not included here) have outcomes very similar to those of Impact 1--4 and can be ruled out as plausible formation scenarios for Arrokoth.

\subsection{Favored scenario}

It is clear that a scenario in which Arrokoth formed when two unbound KBOs collided is highly unlikely. Our simulations demonstrate that even with cohesive strength, such a collision would leave the progenitors either unbound in the case of a grazing collision, or deformed or destroyed in the case of a more direct collision. This holds true even with impact speeds that are small with respect to present-day Kuiper belt crossing speeds of $\sim$300 m/s (Greenstreet \etal, 2019). Even when a recognizable contact binary is formed in this scenario, in the case of the 45$^{\circ}$ impact at 5 m/s, the neck that results is much less well-defined than what we see in images of Arrokoth. This leads us to believe that a history in which LL and SL were in a synchronous, decaying orbit is much more likely.

Most of our simulations had final spin periods shorter than that of Arrokoth. Only the 0.16 g/cm\textsuperscript{3} inspiral scenarios had periods comparable to the observed 15.92 h, and these seem implausible due to the large deformation in the lobes after the merger. This may be an indication that Arrokoth's true density is greater than 0.16 g/cm\textsuperscript{3} and that it initially had a shorter period but continued to lose angular momentum after the merger. The simulations with a bulk density of 0.5 and 0.25 g/cm\textsuperscript{3} accurately reproduce the shape of Arrokoth, but have periods of $\sim9$ h and $\sim12.5$ h respectively---shorter than what is observed.

We can consider our results in the context of the KBO formation mechanisms discussed in Section \ref{S:1}. In the case of hierarchical coagulation (HC), LL and SL would have had to form separately via accretion in different parts of the Kuiper belt before finally coming together and becoming ``stuck" in their current configuration. If they were brought together above the escape speed, the collision would have resulted in disruption of one or both of the lobes, a grazing contact after which the bodies remained unbound, or a grazing contact that resulted in a highly eccentric binary. In the latter case, subsequent collisions between the bodies would probably occur at speeds high enough to damage one or both of the progenitors. Even if a contact binary formed in this way, it would likely not look much like Arrokoth.

To plausibly produce an Arrokoth-like object via HC, the progenitors would need to be brought together after forming separately at speeds $\lesssim$ 3 m/s. This could be accomplished if LL and SL formed a temporary binary system before merging. Collisional binary formation mechanisms like that proposed by Weidenschilling (2002) would not be suitable here, since the speeds involved would be well above the thresholds for destruction of LL and SL that we have described. The exchange reaction scenario described by Funato \etal\ (2004) exclusively produces binaries with $e \gtrsim 0.8$, which leads to the same problem that the grazing collision case has. Goldreich \etal\ (2002) put forward two different collisionless capture pathways for forming binaries in the Kuiper belt. In the first (L$^{2}$s), two bodies become an unstable binary after entering each other's Hill spheres. A background of smaller particles then shrinks the binary and stabilizes it via dynamical friction. In the second (L$^3$), the binary is stabilized by an interaction with a third large body. Once a binary is formed, it must then collapse to the point that the two bodies come into contact. In the Goldreich models, further dynamical friction from surrounding smaller bodies causes the binaries to shrink. For the $\sim$100 km bodies with density 1 gm/cm\textsuperscript{3} considered in the Goldreich paper, the authors estimate that the timescale for a large object to become bound in an equal-mass binary is $3\times10^5$ yr, and that the timescale for the binary to inspiral until the bodies come into contact is 10$^6$ yr (see Eqs. 13 and 14 in that work). In our case (roughly 10 km bodies with density 0.5 g/cm\textsuperscript{3}), the same equations give a binding timescale of $\sim$750 yr and merger timescale of $\sim5\times10^4$ yr. Depending on the lifetime of the Sun's protoplanetary disk, this could explain why larger, 100 km-sized binary systems survived in the Kuiper belt, while Arrokoth's progenitor bodies collapsed to a contact binary. On the other hand, a more recent work (Nesvorn\'{y} \etal, 2019) makes a strong case that the gravitational instability (GI) model produces binary inclinations that match observations much better than both L$^{2}$s and L$^3$ do. Furthermore, the HC model cannot explain the observed color correlation between LL and SL.

The GI pathway for forming KBOs is well-supported by our work. Gravitationally bound concentrations of solid particles form in the midplane of the protosolar disk or as a result of the streaming instability. As they collapse, they could ultimately produce a nearly equal-mass, tightly spaced binary. This binary could then inspiral as it is drained of angular momentum. McKinnon \etal\ (2020) consider a number of possible mechanisms, but point to gas drag as the most convincing. They estimate that due to the pressure gradient in the nebular gas the orbital speed of the gas at Arrokoth's distance from the Sun would be about 1\% of the Keplerian speed, so that a binary system feels a strong ``headwind." This gives rise to torques on both bodies, which could cause the binary to collapse completely over as little as a few million years---within the lifetime of the gas nebula. Gas drag would likely also have a lesser effect on larger bodies, providing a potential explanation for why Arrokoth became a contact binary, but larger bound bodies remained separated. At least for bodies like Arrokoth, GI seems to be the more likely formation pathway. This could also explain why Arrokoth's current spin period is slower than that of the simulated Arrokoth in Inspiral 1--7---it could have continued to lose angular momentum after the merger. And as described in Section \ref{S:3}, a contact binary with the properties of Inspiral 1 can certainly maintain its shape even with the diminished centrifugal support provided by a 16 h spin period.

\section{Conclusions}
\label{S:5}

We used pkdgrav, an N-body tree code with soft-sphere particle interactions, to model the formation of the contact binary (486953) Arrokoth. We focused on modeling the merger itself, and not the origin of the objects that now make up Arrokoth. While images from the New Horizons spacecraft show that Arrokoth is composed of two distinct and apparently primordial lobes connected by a narrow neck, we showed that most direct impacts between the modeled progenitors lead to destruction or significant disruption of the lobes. A more likely formation scenario would involve a Kuiper belt binary that gradually lost angular momentum, spiraled toward a gentle contact to form a contact binary, and continued to spin down, leaving Arrokoth with the $\sim$16 h period that 
we observe today. While KBB formation mechanisms have received a good deal of attention, the literature on the dynamics of binaries in the outer solar system merging to form contact binaries is more sparse. Nesvorn\'{y} and Vokrouhlick\'{y} (2019) find that for cold classical objects of $\sim$100 km, the binary fraction could be as low as 10\%. Thus, if Arrokoth is typical for objects of its size, the fraction of binaries that collapsed to contact must be much higher for smaller objects. This may lend further support to the idea that nebular gas was at least partially responsible for collapsing the Arrokoth binary. A clear next step would be to model an earlier period in Arrokoth's evolution. With an implementation of dynamical friction or gas drag in an N-body code, one could model the inspiral process to determine what initial conditions lead to impacts that could plausibly form an object like Arrokoth, which would give us further insight into planetesimal evolution and planet formation in the outer solar nebula.

\section{Acknowledgements}
This work was supported in part by NASA grant NNX15AH90G awarded by the Solar System Workings program, by a University of Maryland Graduate School Research and Scholarship Award, and by NASA’s New Horizons project via contracts NASW-02008 and NAS5-97271/TaskOrder30. Simulations were performed on the University of Maryland's Deepthought 2 cluster and some visualizations use the Persistence of Vision Ray Tracer (www.povray.org).

\section{References}

\begin{description}

\item \paper {Astakhov, S.A., Lee, E.A., Farrelly, D.} 
  2005 {Formation of Kuiper-belt binaries through multiple 
  chaotic scattering encounters with low-mass intruders} 
  {Mon.\ Not.\ R.\ Astron.\ Soc.} 360 401--415

\item \paper {Batygin, K., Brown, M.E., Fraser, W.C.} 
  2011 {Retention of a primordial cold classical Kuiper 
  belt in an instability-driven model of solar system 
  formation} {Astrophys.\ J.} 738 {article id.\ 13, 8 pp}

\item \paper {Benecchi, S.D., Noll, K.S., Grundy, W.M., 
  Buie, M.W., Stephens, D.C., Levison, H.F.} 2009 {The 
  correlated colors of transneptunian binaries} Icarus 
  200 292--303

\item \paper {Benecchi, S.D., Borncamp, D., Parker, A.H., 
  Buie, M.W., Noll, K.S., Binzel, R.P., Stern, S.A., 
  Verbiscer, A.J., Kavelaars, J.J., Zangari, A.M., 
  Spencer, J.R., Weaver, H.A.} 2019a {The color and binarity 
  of (486958) 2014 MU\textsubscript{69} and other long-range 
  New Horizons Kuiper belt targets} Icarus 334 22--29

\item \paper {Benecchi, S.D., Porter, S.B., Buie, M.W., 
  Zangari, A.M., Verbiscer, A.J., Noll, K.S., Stern, 
  S.A., Spencer, J.R., Parker, A.H.} 2019b {The HST 
  lightcurve of (486958) 2014 MU\textsubscript{69}} Icarus 
  334 11--21

\item \paper {DeMartini, J.V., Richardson, D.C., 
  Barnouin, O.S., Schmerr, N.C., Plescia, J.B., 
  Scheirich, P., Pravec, P.} 2019 {Using a discrete 
  element method to investigate seismic response and
  spin change of 99942 Apophis during its 2029 tidal 
  encounter with Earth} Icarus 328 93--103

\item \paper {Funato, Y., Makino, J., Hut, P., Kokubo, E., 
  Kinoshita, D.} 2004 {The formation of Kuiper-belt binaries 
  through exchange reactions} Nature 427 518--520

\item \paper {Goldreich, P., Lithwick, Y., Sari, R.} 2002 
  {Formation of Kuiper-belt binaries by dynamical friction 
  and three-body encounters} Nature 420 643--646

\item \paper {Greenstreet, S., Gladman, B., McKinnon, W.B., 
  Kavelaars, J.J., Singer, K.N.} 2019 {Crater density predictions 
  for New Horizons flyby target 2014 MU69} {Astrophys.\ J. Lett.} 
  872 {article id.\ L5 6 pp}

\item \paper {Groussin, O., Attree, N., Brouet, Y., 
  Ciarletti, V., Davidsson, B., Filacchione, G., 
  Fischer, H.-H., Gundlach, B., Knapmeyer, M., 
  Knollenberg, J., others} 2019 {The thermal, mechanical, 
  structural, and dielectric properties of cometary 
  nuclei after Rosetta} {Space Sci.\ Rev.} 215 
  {article id.\ 29, 51 pp}

\item \paper {Gulbis, A.A.S., Elliot, J.L., 
  Kane, J.F.} 2006 {The color of the Kuiper 
  belt core} Icarus 183 168--178

\item \paper {Heinisch, P., Auster, H.-U., 
  Gundlach, B., Blum, J., G\"{u}ttler, C., T
  ubiana, C., Sierks, H., Hilchenbach, M., 
  Biele, J., Richter, I., Glassmeier, K.H.} 2019 
  {Compressive strength of comet 67P/Churyumov-Gerasimenko 
  derived from Philae surface contacts} 
  {Astron.\ Astrophys.} 630 {article id.\ A2, 8 pp}

\item \paper {Jorda, L., Gaskell, R., Capanna, C., Hviid, S., 
  Lamy, P., \u{D}urech, J., Faury, G., Groussin, O., 
  Guti\'{e}rrez, P., Jackman, C., others} 2016 {The global shape, 
  density and rotation of comet 67P/Churyumov-Gerasimenko from 
  preperihelion Rosetta/OSIRIS observations} Icarus 277 257--278

\item \paper {Jutzi, M., Asphaug, E.} 2015 {The shape and 
  structure of cometary nuclei as a result of low-velocity 
  accretion} Science 348 1355--1358

\item \paper {Jutzi, M., Benz, W.} 2017 {Formation of bi-lobed 
  shapes by sub-catastrophic collisions. A late origin of comet 
  67P's structure} {Astron.\ Astrophys.} 597 {article id.\ A62, 10 pp}

\item \paper {Jutzi, M.} 2019 {The shape and structure of 
  small asteroids as a result of sub-catastrophic collisions} 
  {Planet.\ Space Sci.} 177 {article id.\ 104695}

\item \paper {Kenyon, S.J.} 2002 {Planet formation in the outer 
  solar system} {Publ.\ Astron.\ Soc.\ Pac.} 114 265--283

\item \paper {Lee, E.A., Astakhov, S.A., Farrelly, D.} 
  2007 {Production of trans-Neptunian binaries through 
  chaos-assisted capture} {Mon.\ Not.\ R.\ Astron.\ Soc.} 
  379 229--246

\item \paper {Leinhardt, Z.M., Richardson, D.C., Quinn, T.} 2000 
  {Direct N-body simulations of rubble pile collisions} Icarus 146 133--151

\item \paper {McKinnon, W.B., Richardson, D.C., Marohnic, J.C., 
   Keane, J.T., Grundy, W.M., Hamilton, D.P., Nesvorn\'{y}, D., 
   Umurhan, O.M., Lauer, T.R., Singer, K.N., others} 2020 
   {The solar nebula origin of (486958) Arrokoth, a primordial 
   contact binary in the Kuiper belt} Science 367 {article id.\ eaay6620}

\item \paper {Nesvorn{\'y}, D., Youdin, A., Richardson, D.C.} 2010 
  {Formation of Kuiper belt binaries by gravitational collapse} 
  {Astron.\ J.} 140 785--793

\item \paper {Nesvorn\'{y}, D., Li, R., Youdin, A.N., Simon, J.B., 
  Grundy, W.M.} 2019 {Trans-Neptunian binaries as evidence for 
  planetesimal formation by the streaming instability} 
  {Nat.\ Astron.} 3 808--812

\item \paper {Nesvorn\'{y}, D., Vokrouhlick\'{y}} 2019 
  {Binary survival in the outer solar system} Icarus 331 49--61

\item \paper {Noll, K.S., Grundy, W.M., Chiang, E.I., 
  Margot, J., Kern, S.D.} 2007 {Binaries in the Kuiper 
  belt} {arXiv preprint} astro-ph 0703134

\item \paper {Noll, K.S., Grundy, W.M., Stephens, D.C., 
  Levison, H.F., Kern, S.D.} 2008 {Evidence for two 
  populations of classical transneptunian objects: the
  strong inclination dependence of classical binaries} 
  Icarus 194 758--768
  
\item \paper {Prialnik, D., Sarid, G., Rosenberg, E.D.,
  Merk, R.} 2008 {Thermal and chemical evolution of comet 
  nuclei and Kuiper belt objects} {Space Sci.\ Rev.} 
  138 147--164

\item \paper {Quillen, A.C., Kueter-Young, A., Frouard, J.,
  Ragozzine, D.} 2016 {Tidal spin down rates of homogeneous 
  triaxial viscoelastic bodies} {Mon.\ Not.\ R.\ Astron.\ Soc.} 
  463 1543--1553

\item \abs {Samarasinha, N.H., Belton, M., Farnham, T., 
  Gutierrez, P., Mueller, B., Chesley, S.} 2009 {Bulk 
  density of comet 9P/Tempel 1 based on orbital and 
  rotational nongravitational effects} {DPS meeting} 
  41 37.05

\item \paper {Schlichting, H.E., Sari, R.} 2008 {Formation 
  of Kuiper belt binaries} {Astrophys.\ J.} 673 1218--1224

\item \paper {Schwartz, S.R., Richardson, D.C., Michel, P.} 2012 {An
  implementation of the soft-sphere discrete element method in a
  high-performance parallel gravity tree-code} {Granul.\ Matter} 14
  363--380

\item \paper {Schwartz, S.R., Michel, P., Jutzi, M., 
  Marchi, S., Zhang, Y., Richardson, D.C.} 2018 
  {Catastrophic disruptions as the origin of bilobate 
  comets} {Nat.\ Astron.} 2 379--382

\item \thesis {Stadel, J.} 2001 {Cosmological $N$-body simulations and
  their analysis} {University of Washington, Seattle} 126

\item \paper {Stern, S.A.} 1996 {On the collision environment, 
  accretion time scales, and architecture of the massive, 
  primordial Kuiper belt} {Astron.\ J} 112 1203--1211

\item \paper {Stern, S.A., Weaver, H.A., Spencer, J.R., 
  Elliott, H.A.} 2018 {The New Horizons Kuiper belt extended 
  mission} {Space Sci.\ Rev.} 214 {article id.\ 77, 23 pp}

\item \paper {Stern, S.A., Weaver, H.A., Spencer, J.R., 
  Olkin, C.B., Gladstone, G.R., Grundy, W.M., Moore, J.M., 
  Cruikshank, D.P., Elliott, H.A., McKinnon, W.B., 
  others} 2019 {Initial results from the New Horizons 
  exploration of 2014 MU\textsubscript{69}} Science 364 
  {article id.\ aaw9771}

\item \paper {Sugiura, K., Kobayashi, H., Inutsuka, S.} 
  2020 {High-resolution simulatiosn of catastrophic 
  disruptions: Resultant shape distributions} 
  {Planet.\ Space Sci.} 181 {article id.\ 104807}

\item \abs {Wandel, O., Wilhelm, K., Christoph, S., 
  Malamud, U., Grishin, E., Perets, H.} 2019 {Numerical 
  simulations on the formation of Ultima Thule} 
  {EPSC-DPS Joint Meeting} 13 1768

\item \paper {Weidenschilling, S.J.} 2002 {On the origin 
  of binary transneptunian objects} Icarus 160 212--215

\item \paper {Zhang, Y., Richardson, D.C., Barnouin, O.S., Maurel, C.,
  Michel, P., Schwartz, S.R., Ballouz, R.-L., Benner, L.A.M., Naidu,
  S.P., Li, J.} 2017 {Creep stability of the proposed AIDA mission
  target 65803 Didymos: I. Discrete cohesionless granular physics
  model} Icarus 294 98--123

\item \paper {Zhang, Y., Richardson, D.C., Barnouin, O.S., 
  Michel, P., Schwartz, S.R., Ballouz, R.-L.} 2018 {Rotational 
  failure of rubble-pile bodies: Influences of shear and 
  cohesive strengths} {Astrophys.\ J.} 857 {article id.\ 15, 20 pp}

\end{description}

\end{document}